\documentclass[sigconf,authorversion,nonacm]{acmart}

\usepackage{graphicx}
\usepackage{multirow}
\usepackage{colortbl}
\usepackage{tcolorbox}
\usepackage{enumerate}
\usepackage{enumitem}
\usepackage{caption}
\usepackage[framemethod=tikz]{mdframed}
\usepackage{float}
\usepackage{placeins}

\usepackage{hyperref}

\usepackage{titlesec}

\mdfdefinestyle{mpdframe}{
    frametitlebackgroundcolor   =black!15,
    frametitlerule              =true,
    roundcorner                 =5pt,
    middlelinewidth             =1pt,
    innermargin                 =0.3cm,
    outermargin                 =0.3cm,
    innerleftmargin             =0.3cm,
    innerrightmargin            =0.3cm,
    innertopmargin              =0.5cm,
    innerbottommargin           =0.5cm
}

\AtBeginDocument{%
  \providecommand\BibTeX{{%
    \normalfont B\kern-0.5em{\scshape i\kern-0.25em b}\kern-0.8em\TeX}}}

\copyrightyear{2024}
\acmYear{2024}
\setcopyright{rightsretained}
\acmConference[MSR '24]{21st International Conference on Mining Software
Repositories}{April 15--16, 2024}{Lisbon, Portugal}
\acmBooktitle{21st International Conference on Mining Software Repositories
(MSR '24), April 15--16, 2024, Lisbon,
Portugal}
\acmDOI{10.1145/3643991.3644929}
\acmISBN{979-8-4007-0587-8/24/04}

\begin{document}

\title{Encoding Version History Context for Better Code Representation}

\author{Huy Nguyen}
\orcid{0000-0002-8796-0762}
\affiliation{%
  \institution{The University of Melbourne}
  \country{Australia}
  \postcode{VIC 3053}
}
\email{huyxuan.nguyen@student.unimelb.edu.au}

\author{Christoph Treude}
\orcid{0000-0002-6919-2149}
\affiliation{%
  \institution{Singapore Management University}
  \country{Singapore}}
\email{ctreude@smu.edu.sg}

\author{Patanamon Thongtanunam}
\orcid{0000-0001-6328-8839}
\affiliation{%
  \institution{The University of Melbourne}
  \country{Australia}}
\email{patanamon.t@unimelb.edu.au}

\renewcommand{\shortauthors}{Nguyen et al.}

\begin{abstract}
  With the exponential growth of AI tools that generate source code, understanding software has become crucial. When developers comprehend a program, they may refer to additional contexts to look for information, e.g. program documentation or historical code versions. Therefore, we argue that encoding this additional contextual information could also benefit code representation for deep learning. Recent papers incorporate contextual data (e.g. call hierarchy) into vector representation to address program comprehension problems. This motivates further studies to explore additional contexts, such as version history, to enhance models' understanding of programs. That is, insights from version history enable recognition of patterns in code evolution over time, recurring issues, and the effectiveness of past solutions. Our paper presents preliminary evidence of the potential benefit of encoding contextual information from the version history to predict code clones and perform code classification. We experiment with two representative deep learning models, ASTNN and CodeBERT, to investigate whether combining additional contexts with different aggregations may benefit downstream activities. The experimental result affirms the positive impact of combining version history into source code representation in all scenarios; however, to ensure the technique performs consistently, we need to conduct a holistic investigation on a larger code base using different combinations of contexts, aggregation, and models. Therefore, we propose a research agenda aimed at exploring various aspects of encoding additional context to improve code representation and its optimal utilisation in specific situations.
\end{abstract}

\ccsdesc[500]{Computing methodologies~Neural networks}
\ccsdesc[500]{Software and its engineering~Reusability}

\keywords{Source code representation, additional context, version history}



\maketitle

\section{Introduction}

Understanding software becomes increasingly crucial to the development and application of technology to satisfy user demands~\cite{gold2004understanding}. 
Understanding complex software systems is often challenging. These challenges are compounded by time constraints and constantly changing business requirements~\cite{sites2021understanding}. 
Artificial Intelligence (AI) can help developers generate code quickly and efficiently~\cite{bano2024large, pornprasit2023d, gitclear2023CodingCopilot}, contributing to the growth of new source code in various domains, e.g. education~\cite{eke2023chatgpt, pan2024assessing}. 
We argue that understanding software has become more essential than ever. However, the effectiveness of these AI tools strongly depends on their ability to comprehend the given context and the generated source code~\cite{nijkamp2022codegen}. 
Therefore, improving AI's ability to understand source code and contextual information is critical to ensure that the outputs of these tools are reliable~\cite{yang2023learning}.
Although recent studies have advanced source code representation, they also reveal significant research gaps. Recent articles have predominantly focused on harnessing deep learning techniques for software comprehension tasks. However, they have often neglected the full utilisation of additional contexts that can significantly improve performance~\cite{wang2023comparison}. 
Although these studies shed light on various aspects, they often rely on relatively old and simplistic datasets, e.g. OnlineJudge or BigCloneBench~\cite{samoaa2022systematic}. 
These datasets are reliable regarding the volume or annotation of data, but they are limited in terms of additional contexts. Since the popularity of code hosting platforms, e.g. GitHub, crawling other contextual information has become easier~\cite{tian2022adding}.
For example, Wang and Lo~\cite{wang2014version} claim that putting together version history, similar reports, and structure can help locate relevant buggy files. 
However, the proposed statistical method uses only the latest version of the code and does not incorporate it into the code representation for downstream tasks. We argue that deep learning models may also comprehend source code better if they can access information beyond source code (e.g. version history). 


\begin{figure*}[h]
  \centering
  \includegraphics[width=0.95\linewidth]{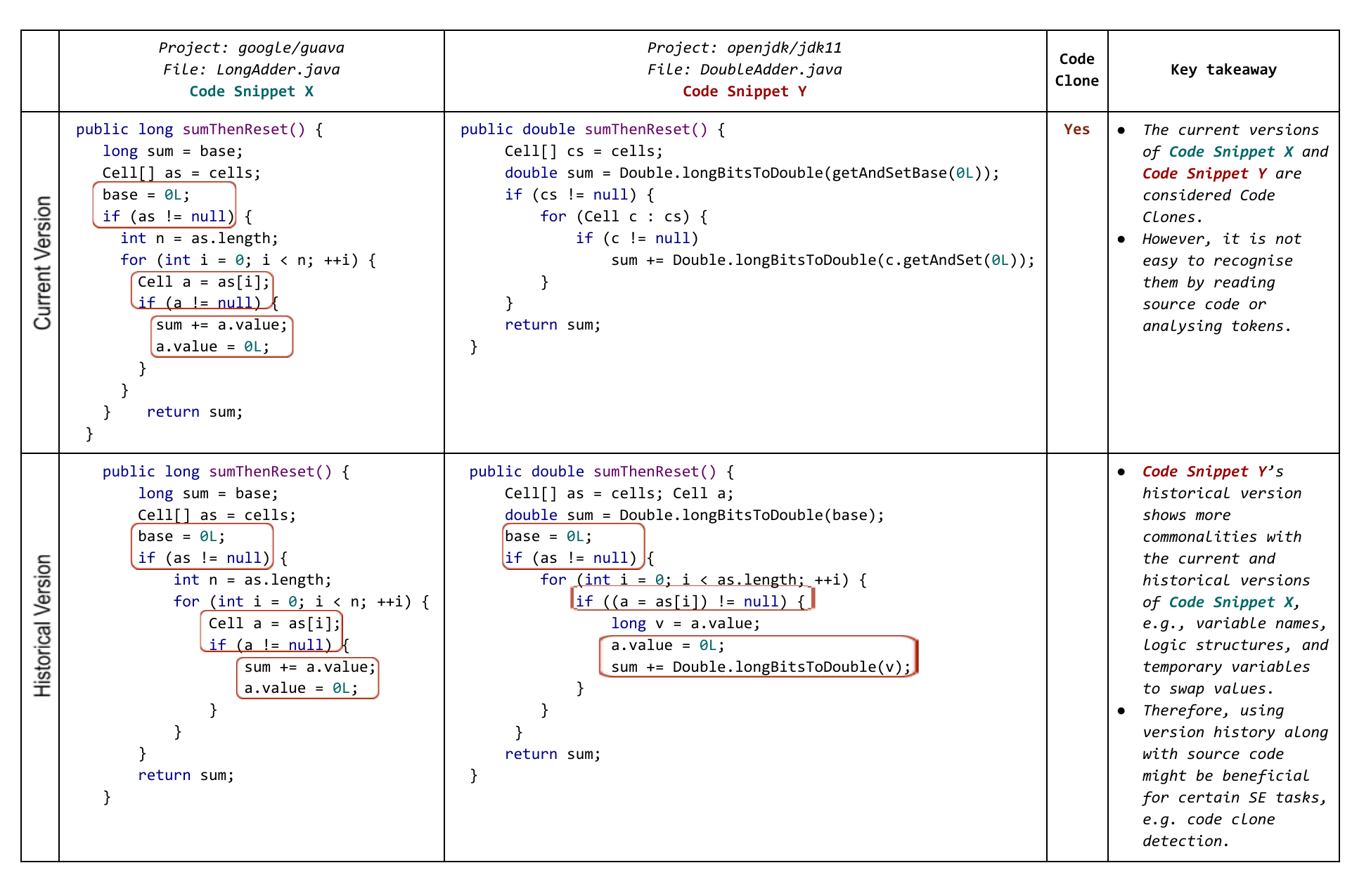}
  \caption{A motivating example of using Version History to detect code clones.}
  \label{fig:code_example}
\end{figure*}

Figure \ref{fig:code_example} illustrates a motivating example of a code clone pair of two Java methods and their historical versions from two open-source projects on GitHub.
Their historical versions can help recognise code clones since their history shows more commonalities than their current versions~\footnote{\href{https://github.com/openjdk/jdk11/blob/00d1900dc994bb745654c93bdca95ad3e579f720/src/java.base/share/classes/java/util/concurrent/atomic/DoubleAdder.java\#L155}{project: jdk11 | file: DoubleAdder.java | method: sumThenReset() | current version}}~\footnote{\href{https://github.com/openjdk/jdk11/blob/3f14786363e002a7f9f01713349a4ad687677f11/jdk/src/share/classes/java/util/concurrent/atomic/DoubleAdder.java\#L156}{project: jdk11 | file: DoubleAdder.java | method: sumThenReset() | historical version}}~\footnote{\href{https://github.com/google/guava/blob/fd919e54a55ba169dc7d9f54b7b3485aa7fa0970/android/guava/src/com/google/common/cache/LongAdder.java\#L123}{project: guava | file: LongAdder.java | method: sumThenReset() | current version}}~\footnote{\href{https://github.com/google/guava/blob/06cbdb1ed84f235b51fec6acc740d8261357545d/guava/src/com/google/common/cache/LongAdder.java\#L135}{project: guava | file: LongAdder.java | method: sumThenReset() | historical version}}.
The takeaway from the example is that using version history along with source code could be beneficial for certain SE tasks, e.g. code clone detection or bug localisation.

We present the initial results of encoding version history context for better code representation into two representative deep learning models, i.e. ASTNN~\cite{zhang2019novel, tian2022adding} and CodeBERT~\cite{feng2020codebert}. Experimental results on two software engineering tasks, i.e. Code Clone Detection and Code Classification, indicate that adding multiple historical versions of code to the final representation improves the models' performance compared to using only the original code snippet. 

Our experimental results show that concatenating only version history information can boost the model performance (F1 score) by 15\% (from 0.667 to 0.769) for CodeBERT and by 7\% (from 0.824 to 0.880) for ASTNN. We can achieve even better results when using multiple additional contexts (version history, call hierarchy, and the number of existing days). Namely, CodeBERT's F1 score for Code Clone Detection increases by 27\% to 0.846 when we concatenate the representation of the absolute difference between the two methods with both version history context and number of days information. However, we also observe poorer performance in some scenarios. 

We conclude that the version history context can improve the code representation for deep learning, but how to best use it requires further investigation. 
From these findings, we propose our research agenda to explore various aspects of encoding additional contexts, especially version history, to improve code representation and its optimal utilisation in specific situations. 



\section{Related Work}

Our research combines knowledge from two aspects, i.e. source code representation and program comprehension. 

\textbf{Source Code Representation.} Source code, written by programmers or generated by tools, is initially a text-encoded representation of a program. Therefore, it can be converted into various forms of representation. An effective code representation could benefit program comprehension tasks, such as program repair or code clone detection~\cite{namavar2022controlled}. 

Determining the appropriate representation of source code is thus a crucial aspect of many software engineering tasks. Recent papers have introduced popular techniques for addressing different downstream tasks, including graph-based, tree-based, or token-based techniques. 
In 2019, a well-known neural-based code representation for code, called ASTNN, used tree-based CNN to transform AST sub-trees into vector format~\cite{zhang2019novel,tian2022adding}. Many other approaches use tree- or graph-based representations for bug detection or program classification~\cite{guo2020graphcodebert,li2019improving,zhang2023implant}.


Furthermore, existing studies on transformer-based models for programming languages use a tokeniser to convert the text input to numerical representation that can be processed by the model~\cite{feng2020codebert}. Other studies performed deep learning tasks using code representation, incorporating high-level semantic and low-level syntactic information~\cite{jiang2022hierarchical}. Hybrid representation techniques are becoming increasingly popular, where more than one code representation can be used~\cite{samoaa2022systematic,tian2022adding,long2022multi}.

However, there is an existing research gap on how to improve the input of representation techniques. We argue that many available programming artefacts, e.g. version history, that go beyond source code could benefit code representation and downstream tasks.

\textbf{Context Considered by Humans During Program Comprehension.} Understanding software is a term in software engineering research that encompasses both the human and deep learning perspective~\cite{sites2021understanding}. We argue that additional contexts from the software development process support developers in comprehending source code; therefore, it may work similarly for deep learning.

Maletic and Marcus~\cite{maletic2001supporting} claim that multiple software artefacts, with semantic and structural context, provide valuable support for program comprehension. Furthermore, Kulkarni and Varma~\cite{kulkarni2014supporting} indicated that the cues derived from different programming contexts help establish the relevance of information for software engineering tasks. In addition, developers may be interested in task-related software artefacts to understand program logic rather than using the entire source code~\cite{sharif2016tracking}. 

Most recent studies on deep learning only use the source code itself as an input~\cite{wang2023comparison, long2022multi}. However, few papers focus on exploring different types of data~\cite{hoang2020cc2vec, tian2022adding, jiang2022hierarchical, pornprasit2023d}, for instance, version history or execution traces. The main reason could be the challenges of mining data from artefacts produced as part of the software engineering process or the limited computational resources to train the models~\cite{samoaa2022systematic}. However, the fast evolution of foundation models, such as Large Language Models (LLMs) like GPT-4, could help to overcome current challenges in proposing new code representation techniques~\cite{bano2024large, pornprasit2023d, eke2023chatgpt, pan2024assessing}.

\section{Preliminary Study}


This section outlines our preliminary study approach to investigate the feasibility of encoding version history to source code representation and how it improves deep learning's performance in software engineering tasks. 
To conduct the study, we 1) mine a version history in a code repository, 2) explore suitable aggregation techniques, and 3) evaluate the performance of two well-known models (ASTNN and CodeBERT) on downstream tasks.
Lastly, we explore combinations of multiple contexts and their effectiveness.



To evaluate the benefits of adding version history to source code representation, we set out the following three research questions:
\begin{enumerate}[label=\textbf{RQ\arabic*},leftmargin=*]
    \item  What is the impact of encoding of version history on the performance of deep learning models?

    \item  What is the impact of different aggregation techniques on the representation of source code and its version history context?
    
    \item  How does combining multiple additional contexts into source code representation impact deep learning models?
\end{enumerate}


\paragraph{Data Collection.}

Since the version history we need is only available on the source hosting system, we need a dataset that contains the repository information. Hence, we use SeSaMe~\cite{kamp2019sesame}, a dataset of semantically similar Java methods from 11 software projects, all available on code hosting platforms (e.g. GitHub). The version history context refers to all changes to a method (or a code fragment) during its lifetime, and each version is a particular snapshot. Thus, every method has a version history.

To extract version history data, we use \textit{PyDriller}~\cite{spadini2018pydriller} to walk through all commits based on the provided \textit{commit hash}. We also use \textit{Lizard} to analyse the source code and extract only the methods from the SeSaMe dataset~\cite{thongtanunam2019will}. Lastly, we keep only versions in which the method's source code was changed.
Along with version history, we extract call hierarchy (caller and callee)~\cite{tian2022adding} and number of days as additional contexts for the experiment. The number of days contains a numerical value that describes how long a method existed in the repository. We argue that this numerical information might indicate some relationship between the two methods and help to detect code clones.

Table ~\ref{tab:tbl_stats_analysis} above introduces descriptive statistics of our dataset. We extracted 10,531 code versions of 1,679 unique Java methods from 11 open-source projects.
The number of methods per project and the number of versions per method are diverse, ranging from an average of 1 version/method (\textit{trove}) to an average of 26.71 versions/method (\textit{checkstyle}). A method's lifetime varies from 17 to 6,334 days, and the average number of changed lines/version is 3.94 lines/version.

\begin{table}[]
\caption{Statistical Analysis of the Dataset}
\label{tab:tbl_stats_analysis}
\resizebox{\columnwidth}{!}{%
\begin{tabular}{lrrrc}
\hline
\textbf{GitHub Project} &
  \multicolumn{1}{c}{\textbf{\begin{tabular}[c]{@{}c@{}}\# of \\ methods\end{tabular}}} &
  \multicolumn{1}{c}{\textbf{\begin{tabular}[c]{@{}c@{}}Avg \# of\\ version\\ /method\end{tabular}}} &
  \multicolumn{1}{c}{\textbf{\begin{tabular}[c]{@{}c@{}}Avg \# of\\ changed \\ lines\\ /version\end{tabular}}} &
  \textbf{\begin{tabular}[c]{@{}c@{}}Min|Max|Avg\\ \# of days\end{tabular}} \\ \hline
{\color[HTML]{333333} \textit{caffeine}} &
  {\color[HTML]{333333} 63} &
  {\color[HTML]{333333} 2.25} &
  {\color[HTML]{333333} 1.14} &
  {\color[HTML]{333333} 196 | 1,328 | 1,174} \\
{\color[HTML]{333333} \textit{checkstyle}} &
  {\color[HTML]{333333} 52} &
  {\color[HTML]{333333} 26.71} &
  {\color[HTML]{333333} 3.57} &
  {\color[HTML]{333333} 125 | 1,665 |    989} \\
{\color[HTML]{333333} \textit{commons-collections}} &
  {\color[HTML]{333333} 81} &
  {\color[HTML]{333333} 1.51} &
  {\color[HTML]{333333} 0.56} &
  {\color[HTML]{333333} 273 | 1,994 | 1,883} \\
{\color[HTML]{333333} \textit{commons-lang}} &
  {\color[HTML]{333333} 57} &
  {\color[HTML]{333333} 9.51} &
  {\color[HTML]{333333} 1.76} &
  {\color[HTML]{333333} 535 | 3,198 | 2,815} \\
{\color[HTML]{333333} \textit{commons-math}} &
  {\color[HTML]{333333} 93} &
  {\color[HTML]{333333} 1.51} &
  {\color[HTML]{333333} 3.51} &
  {\color[HTML]{333333} 474 | 1,290 | 1,234} \\
{\color[HTML]{333333} \textit{deeplearning4j}} &
  {\color[HTML]{333333} 212} &
  {\color[HTML]{333333} 1.39} &
  {\color[HTML]{333333} 1.44} &
  {\color[HTML]{333333} 17 |    140 |  136} \\
{\color[HTML]{333333} \textit{eclipse.jdt.core}} &
  {\color[HTML]{333333} 178} &
  {\color[HTML]{333333} 22.80} &
  {\color[HTML]{333333} 4.41} &
  {\color[HTML]{333333} 200 | 6,334 | 4,317} \\
{\color[HTML]{333333} \textit{freemind}} &
  {\color[HTML]{333333} 87} &
  {\color[HTML]{333333} 2.29} &
  {\color[HTML]{333333} 4.71} &
  {\color[HTML]{333333} 299 | 2,742 | 2,335} \\
{\color[HTML]{333333} \textit{guava}} &
  {\color[HTML]{333333} 156} &
  {\color[HTML]{333333} 3.96} &
  {\color[HTML]{333333} 2.53} &
  {\color[HTML]{333333} 378 | 2,720 | 2,184} \\
{\color[HTML]{333333} \textit{openjdk11}} &
  {\color[HTML]{333333} 688} &
  {\color[HTML]{333333} 4.38} &
  {\color[HTML]{333333} 4.65} &
  {\color[HTML]{333333} 58 |    413 |  335} \\
{\color[HTML]{333333} \textit{trove}} &
  {\color[HTML]{333333} 12} &
  {\color[HTML]{333333} 1.00} &
  {\color[HTML]{333333} -} &
  {\color[HTML]{333333} 1,593   | 1,593 | 1,593} \\ \hline
\end{tabular}%
}
\end{table}

\paragraph{Encoding and Aggregation.}

We selected two popular model architectures, ASTNN and CodeBERT, to evaluate the impact of encoding version history context into source code representation in Code Clone detection.
ASTNN uses tree-based architecture, allowing the model to capture hierarchical structural information based on its understanding of source code patterns~\cite{tian2022adding, zhang2019novel}. CodeBERT is constructed from a bimodal pre-trained model using six programming languages~\cite{feng2020codebert, zeng2022extensive}. Unlike ASTNN, taking input as ASTs, CodeBERT accepts code snippets.


To explore different combinations of contexts to observe their interaction within code representation, we design our model experiment into three steps, including (1) Encoding, (2) Aggregation, and (3) Model Training.


First, we use the corresponding technique from ASTNN and CodeBERT to convert the method's source code and its additional contexts into vector representation ~\cite{zhang2019novel, feng2020codebert}. The output derived from a method's source code is a single vector, and the output derived from version history contains a long vector that consists of multiple vectors parsed from historical versions~\cite{tian2022adding, zeng2022extensive}. We follow a recent study~\cite{tian2022adding} to select the longest caller and callee from the call hierarchy to produce two separate vectors for caller and callee, respectively. Also, the number of days is also transformed into a vector. After the Encoding step, we have five vectors (or numerical representations) in total representing information of the method's current source code, multiple historical versions, caller, callee, and number of days.


Secondly, to combine selections of vector representations from the previous step, we select three aggregation methods: 1) concatenation, 2) max-pooling, and 3) concatenation of absolute difference. These are the aggregation methods that are suitable for both vector representations (for ASTNN) and text representations (for CodeBERT)~\cite{gholamalinezhad2020pooling}. In our preliminary study, we select these three methods since they are well-established approaches within our problem domain~\cite{tian2022adding}. Finally, we pass the output from the Aggregation step into a linear layer and a sigmoid layer (as the Model training step) to determine whether the two methods are code clones.



\textit{Concatenation} refers to merging the representation of source code with representations of its additional contexts~\cite{gholamalinezhad2020pooling}. We compute the absolute value of the difference between concatenated vectors and pass it into a linear layer to predict cloned code.


\textit{Max-pooling} refers to the pooling technique in deep neural networks~\cite{gholamalinezhad2020pooling}, where we select a vector with the highest values in each dimension among all input vectors from two methods, composed of the method source code and the version history context. Then, we pass it to a linear layer. 


\textit{Concatenation of absolute difference} computes the difference between vectors of two methods' source code first, then merges the output vector with all remaining additional context vectors before passing it to the linear layer function~\cite{tian2022adding}.


Please refer to our online appendix~\cite{package} for all three aggregation scenarios that take the above representation as input.


In Code Classification, we only have concatenation and max-pooling scenarios because the input contains only one method. After aggregating the vectors for the respective scenarios, we pass them to a softmax layer for a classifier, where the output is an array of probabilities for each label in the dataset.

\paragraph{Experimental Setup.}

The SeSaMe dataset allows us to experiment with two software engineering tasks~\cite{zhang2019novel}: 

\textit{Code Clone Detection:} We compute the label from human annotation data on code clone pairs in the SeSaMe dataset~\cite{tian2022adding}. Each pair of codes contains a binary label (0/1), which was constructed based on weights to reflect high, medium, and low confidence. The evaluation metric for this task is the F1 score.  

\textit{Code Classification:} The original dataset contains the 11 GitHub project names associated with code snippets. We use this information as labels for the classification task. For the classification task, we use Accuracy as the evaluation metric.

\paragraph{Training settings.} To ensure fair comparison of our experiment with baseline performance, we also adopt training, validation, and testing sets with an 80:10:10 ratio.
We also use the same hyper-parameters settings, workflows and loss functions for all models. We selected the models with the best results on the validation set.

\section{Experimental Results}
Table \ref{tab:tbl_code_clone} and \ref{tab:tbl_code_classification} present the performance of ASTNN and CodeBERT when combining version history with source code representation for Code Clone Detection and Code Classification tasks, respectively.
We now present results to answer each research question.

\textbf{RQ1: Impact of Adding Version History to Code Representation.}
\begin{table}[]
\caption{Code Clone Detection using Version History or Multiple Contexts}
\label{tab:tbl_code_clone}
\resizebox{\columnwidth}{!}{%
\begin{tabular}{clllrrrr}
\hline
\textbf{} &
   &
  \multicolumn{1}{c}{\textbf{Context(s)}} &
  \multicolumn{1}{c}{\textbf{Aggregation}} &
  \multicolumn{1}{c}{\textbf{P}} &
  \multicolumn{1}{c}{\textbf{R}} &
  \multicolumn{1}{c}{\textbf{F1}} &
  \multicolumn{1}{c}{\textbf{\%F1$\uparrow$}} \\ \hline
\rowcolor[HTML]{EFEFEF} 
\cellcolor[HTML]{C0C0C0} &
   &
  Without Context &
   &
  0.913 &
  0.750 &
  \textbf{0.824} &
   \\ \cline{2-8} 
\cellcolor[HTML]{C0C0C0} &
  * &
  Version History &
  Concatenation &
  1.000 &
  0.786 &
  \textbf{0.880} &
  \textbf{7\%} \\
\cellcolor[HTML]{C0C0C0} &
   &
   &
  Max-pooling &
  0.821 &
  0.821 &
  0.821 &
  0\% \\
\cellcolor[HTML]{C0C0C0} &
   &
   &
  Diff \& Concat &
  0.833 &
  0.714 &
  0.769 &
  -6\% \\ \cline{2-8} 
\cellcolor[HTML]{C0C0C0} &
  * &
  Call Hierarchy &
  Concatenation &
  0.913 &
  0.750 &
  0.824 &
  0\% \\
\cellcolor[HTML]{C0C0C0} &
   &
   &
  Max-pooling &
  0.885 &
  0.821 &
  0.852 &
  4\% \\
\cellcolor[HTML]{C0C0C0} &
   &
   &
  Diff \& Concat &
  0.955 &
  0.750 &
  0.840 &
  2\% \\ \cline{2-8} 
\cellcolor[HTML]{C0C0C0} &
  ** &
  Version History &
  Concatenation &
  0.885 &
  0.821 &
  0.852 &
  4\% \\
\cellcolor[HTML]{C0C0C0} &
   &
  + Call Hierarchy &
  Max-pooling &
  0.852 &
  0.821 &
  0.836 &
  2\% \\
\cellcolor[HTML]{C0C0C0} &
   &
   &
  Diff \& Concat &
  0.875 &
  0.750 &
  0.808 &
  -2\% \\ \cline{2-8} 
\cellcolor[HTML]{C0C0C0} &
  ** &
  Version History &
  Concatenation &
  1.000 &
  0.786 &
  \textbf{0.880} &
  \textbf{7\%} \\
\cellcolor[HTML]{C0C0C0} &
   &
  + No. of Days &
  Max-pooling &
  0.852 &
  0.821 &
  0.836 &
  2\% \\
\cellcolor[HTML]{C0C0C0} &
   &
   &
  Diff \& Concat &
  0.913 &
  0.750 &
  0.824 &
  0\% \\ \cline{2-8} 
\cellcolor[HTML]{C0C0C0} &
  ** &
  Version History &
  Concatenation &
  0.885 &
  0.821 &
  0.852 &
  4\% \\
\cellcolor[HTML]{C0C0C0} &
   &
  + Call Hierarchy &
  Max-pooling &
  0.920 &
  0.821 &
  0.868 &
  6\% \\
\multirow{-16}{*}{\cellcolor[HTML]{C0C0C0}\textbf{
\parbox[t]{2mm}{\multirow{-2}{*}{\rotatebox[origin=c]{90}{ASTNN}}}
}} &
   &
  + No. of Days &
  Diff \& Concat &
  0.864 &
  0.679 &
  0.760 &
  -7\% \\
\multicolumn{1}{l}{} &
   &
   &
   &
  \multicolumn{1}{l}{} &
  \multicolumn{1}{l}{} &
  \multicolumn{1}{l}{} &
  \multicolumn{1}{l}{} \\ \hline
\rowcolor[HTML]{EFEFEF} 
\cellcolor[HTML]{C0C0C0} &
   &
  Without Context &
   &
  0.655 &
  0.679 &
  \textbf{0.667} &
   \\ \cline{2-8} 
\cellcolor[HTML]{C0C0C0} &
  * &
  Version History &
  Concatenation &
  0.778 &
  0.750 &
  0.764 &
  15\% \\
\cellcolor[HTML]{C0C0C0} &
   &
   &
  Max-pooling &
  0.714 &
  0.714 &
  0.714 &
  7\% \\
\cellcolor[HTML]{C0C0C0} &
   &
   &
  Diff \& Concat &
  0.833 &
  0.714 &
  \textbf{0.769} &
  \textbf{15\%} \\ \cline{2-8} 
\cellcolor[HTML]{C0C0C0} &
  * &
  Call Hierarchy &
  Concatenation &
  0.821 &
  0.821 &
  0.821 &
  23\% \\
\cellcolor[HTML]{C0C0C0} &
   &
   &
  Max-pooling &
  0.864 &
  0.679 &
  0.760 &
  14\% \\
\cellcolor[HTML]{C0C0C0} &
   &
   &
  Diff \& Concat &
  0.840 &
  0.750 &
  0.792 &
  19\% \\ \cline{2-8} 
\cellcolor[HTML]{C0C0C0} &
  ** &
  Version History &
  Concatenation &
  0.840 &
  0.750 &
  0.792 &
  19\% \\
\cellcolor[HTML]{C0C0C0} &
   &
  + Call Hierarchy &
  Max-pooling &
  0.800 &
  0.714 &
  0.755 &
  13\% \\
\cellcolor[HTML]{C0C0C0} &
   &
   &
  Diff \& Concat &
  0.815 &
  0.786 &
  0.800 &
  20\% \\ \cline{2-8} 
\cellcolor[HTML]{C0C0C0} &
  ** &
  Version History &
  Concatenation &
  0.679 &
  0.679 &
  0.679 &
  2\% \\
\cellcolor[HTML]{C0C0C0} &
   &
  + No. of Days &
  Max-pooling &
  0.643 &
  0.643 &
  0.643 &
  -4\% \\
\cellcolor[HTML]{C0C0C0} &
   &
   &
  Diff \& Concat &
  0.917 &
  0.786 &
  \textbf{0.846} &
  \textbf{27\%} \\ \cline{2-8} 
\cellcolor[HTML]{C0C0C0} &
  ** &
  Version History &
  Concatenation &
  0.875 &
  0.750 &
  0.808 &
  21\% \\
\cellcolor[HTML]{C0C0C0} &
   &
  + Call Hierarchy &
  Max-pooling &
  0.857 &
  0.643 &
  0.735 &
  10\% \\
\multirow{-16}{*}{\cellcolor[HTML]{C0C0C0}\textbf{
\parbox[t]{2mm}{\multirow{-2}{*}{\rotatebox[origin=c]{90}{CodeBERT}}}
}} &
   &
  + No. of Days &
  Diff \& Concat &
  0.846 &
  0.786 &
  0.815 &
  22\% \\
   \hline
\multicolumn{1}{l}{} &
   &
  \multicolumn{6}{r}{*: Single Context | **: Multiple Contexts}
\end{tabular}%
}
\end{table}
We compare the ASTNN's and CodeBERT's performance between (i) without additional context (baseline) and (ii) with version history using the concatenation aggregation. 

Table \ref{tab:tbl_code_clone} 
shows that for the Code Clone Detection task, the F1 score of the ASTNN and CodeBERT models with version history (using concatenation aggregation) 
increases by 7\% and 15\% compared to the models without additional context.
Similarly, Table \ref{tab:tbl_code_classification} shows that for the Code Classification task, the accuracy of the ASTNN and CodeBERT models with version history increases by 6\% and 4\%, respectively.
These results suggest that encoding version history context, which contains multiple source code versions, helps deep learning perform better than without context.

\textbf{RQ2: Impact of Different Aggregation Techniques.} 
We select the experiment of encoding version history context to source code representation in both software engineering tasks. Among the proposed scenarios, no technique is always better than others in all models and tasks. 

In Code Clone Detection, ASTNN with the concatenation of version history to code representation achieves the highest F1 score of 0.880 (7\% increase compared to the baseline); nevertheless, both concatenation scenarios in CodeBERT achieve 15\% improvement. The max-pooling scenario gains the lowest improvement in both models, only 7\% with CodeBERT and even 0\% with ASTNN. On the contrary, the experiment with the Code Classification task displays another tendency. CodeBERT with max-pooling scenarios achieved a 7\% improvement (0.852 in accuracy). In ASTNN, concatenating techniques increase accuracy by 6\%, from 0.583 (baseline) to 0.617. 

The reason why no aggregation technique outperforms others in all experiments can be explained by how we handle multiple historical code versions to create the final representation. Each method may have one or hundreds of versions during its lifetime. Concatenation and Concatenation of Absolute Difference merge vector representations and then rely on the model's learning capability. ASTNN and CodeBERT have limitations on the input length; therefore, a method with too many versions may be truncated. On the other hand, max-pooling relies on selecting the maximum value within the pool of vector representations. Therefore, some critical information in the unselected representation may be dismissed.



\textbf{RQ3: Impact of Adding Multiple Artefacts to Code Representation.} In this section, we aim to answer RQ3 on the impact of combining multiple programming artefacts, i.e. version history with other contextual information (refer to \textbf{**} in Tables \ref{tab:tbl_code_clone} and \ref{tab:tbl_code_classification}). 



In Clone Detection, combining version history, number of days, and method's source code to new code representation achieves the highest F1 score. Namely, ASTNN with concatenation scenario achieves 0.88 in the F1 score and 7\% improvement. In addition, CodeBERT, with the concatenation of absolute differences, improves the F1 score by 27\% compared to the baseline, equivalent to 0.846. The Code Classification result shows that the combination of version history and call hierarchy context achieves the highest accuracy, increasing by 27\% to 0.739 in ASTNN with concatenation scenario. CodeBERT with max-pooling improves accuracy by 12\%, from 0.645 to 0.896 in Code Classification problems.

Our technique for encoding the version history to source code representation is in its infancy, and the dataset is modest in terms of size and diversity of code base. This may explain why the experiment results do not show a stable tendency. However, if we select the suitable model and aggregation methods, the technique may improve by up to 27\% compared to baseline performance without context. Accordingly, we may obtain the best result if we combine suitable additional artefacts with appropriate techniques.



\begin{table}[]
\caption{Code Classification with Version History or Multiple Contexts}
\label{tab:tbl_code_classification}
\resizebox{\columnwidth}{!}{%
\begin{tabular}{clllrrrr}
\hline
 &
   &
  \multicolumn{1}{c}{\textbf{Context(s)}} &
  \multicolumn{1}{c}{\textbf{Aggregation}} &
  \multicolumn{1}{c}{\textbf{Acc}} &
  \multicolumn{1}{c}{\textbf{P}} &
  \multicolumn{1}{c}{\textbf{R}} &
  \multicolumn{1}{c}{\textbf{\%Acc$\uparrow$}} \\ \hline
\rowcolor[HTML]{EFEFEF} 
\cellcolor[HTML]{C0C0C0} &    & Without Context                                             &               & \textbf{0.583}          & 0.515 & 0.414 &               \\ \cline{2-8} 
\cellcolor[HTML]{C0C0C0} & *  & Version History                                             & Concatenation & \textbf{0.617} & 0.557 & 0.471 & \textbf{6\%}  \\
\cellcolor[HTML]{C0C0C0} &    &                                                             & Max-pooling   & 0.591          & 0.563 & 0.481 & 1\%           \\ \cline{2-8} 
\cellcolor[HTML]{C0C0C0} & *  & Call Hierarchy                                              & Concatenation & 0.713          & 0.640 & 0.558 & 22\%          \\
\cellcolor[HTML]{C0C0C0} &    &                                                             & Max-pooling   & 0.652          & 0.623 & 0.547 & 12\%          \\ \cline{2-8} 
\cellcolor[HTML]{C0C0C0} & ** & Version History                                             & Concatenation & \textbf{0.739} & 0.705 & 0.588 & \textbf{27\%} \\
\cellcolor[HTML]{C0C0C0} &    & + Call Hierarchy                                            & Max-pooling   & 0.704          & 0.739 & 0.554 & 21\%          \\ \cline{2-8} 
\cellcolor[HTML]{C0C0C0} & ** &                                                             & Concatenation & 0.600          & 0.568 & 0.459 & 3\%           \\
\cellcolor[HTML]{C0C0C0} &
   &
  \multirow{-2}{*}{\begin{tabular}[c]{@{}l@{}}Version History\\    + No. of Days\end{tabular}} &
  Max-pooling &
  0.678 &
  0.710 &
  0.545 &
  16\% \\ \cline{2-8} 
\cellcolor[HTML]{C0C0C0} & ** &                                                             & Concatenation & 0.722          & 0.688 & 0.608 & 24\%          \\
\multirow{-11}{*}{\cellcolor[HTML]{C0C0C0}\textbf{
\parbox[t]{2mm}{\multirow{-2}{*}{\rotatebox[origin=c]{90}{ASTNN}}}
}} &
   &
  \multirow{-2}{*}{\begin{tabular}[c]{@{}l@{}}Version History\\    + Call Hierarchy \\    + No. of Days\end{tabular}} &
  Max-pooling &
  0.687 &
  0.663 &
  0.528 &
  18\% \\
                         &    &                                                             &               &                &       &       &               \\ \hline
\rowcolor[HTML]{EFEFEF} 
\cellcolor[HTML]{C0C0C0} &    & \multicolumn{1}{c}{\cellcolor[HTML]{EFEFEF}Without Context} &               & \textbf{0.800}          & 0.753 & 0.645 &               \\ \cline{2-8}
\cellcolor[HTML]{C0C0C0} & *  & Version History                                             & Concatenation & 0.835          & 0.851 & 0.693 & 4\%           \\
\cellcolor[HTML]{C0C0C0} &    &                                                             & Max-pooling   & \textbf{0.852} & 0.777 & 0.772 & \textbf{7\%}  \\ \cline{2-8} 
\cellcolor[HTML]{C0C0C0} & *  & Call Hierarchy                                              & Concatenation & 0.843          & 0.845 & 0.700 & 5\%           \\
\cellcolor[HTML]{C0C0C0} &    &                                                             & Max-pooling   & 0.835          & 0.830 & 0.706 & 4\%           \\ \cline{2-8} 
\cellcolor[HTML]{C0C0C0} & ** & Version History                                             & Concatenation & 0.817          & 0.767 & 0.674 & 2\%           \\
\cellcolor[HTML]{C0C0C0} &    & + Call Hierarchy                                            & Max-pooling   & \textbf{0.896} & 0.847 & 0.812 & \textbf{12\%} \\ \cline{2-8} 
\cellcolor[HTML]{C0C0C0} & ** & Version History                                             & Concatenation & 0.826          & 0.806 & 0.706 & 3\%           \\
\cellcolor[HTML]{C0C0C0} &    & + No. of Days                                               & Max-pooling   & 0.835          & 0.787 & 0.700 & 4\%           \\ \cline{2-8} 
\cellcolor[HTML]{C0C0C0} & ** &                                                             & Concatenation & 0.870          & 0.862 & 0.773 & 9\%           \\
\multirow{-11}{*}{\cellcolor[HTML]{C0C0C0}\textbf{
\parbox[t]{2mm}{\multirow{-2}{*}{\rotatebox[origin=c]{90}{CodeBERT}}}
}} &
   &
  \multirow{-2}{*}{\begin{tabular}[c]{@{}l@{}}Version History\\    + Call Hierarchy \\    + No. of Days\end{tabular}} &
  Max-pooling &
  0.817 &
  0.829 &
  0.638 &
  2\% \\
\multicolumn{1}{l}{}     &    &                                                             &               &                &       &       &               \\ \hline
\multicolumn{1}{l}{}     &    & \multicolumn{6}{r}{*: Single Context | **: Multiple Contexts}                                                               
\end{tabular}%
}
\vspace{-0.5cm}

\end{table}

\section{Threats to Validity and Limitations}

We now discuss possible threats to the validity of the results and limitations.
%
First, as our final vector presentation is concatenated from all historical versions, the vector may be truncated due to CodeBERT's maximum length of input sequences (512 tokens).
In our dataset, the total number of tokens of a method varies from 38 to 369,824 tokens.
We observed only 38\% (637/1,679) of the methods might be impacted by the truncation issue. 
We arranged all versions from the most recent to the oldest version and exhaustively concatenated tokens until they reached the model’s limit.

Second, our experiment is based on a single dataset. Thus, a statistical test for performance improvement is not applicable.
Nevertheless, we quantify the improvement by measuring the percentage difference in F1-score (for Code Clone Detection) and Accuracy (for Code Classification) between adding version history (and other contexts) against the `without-context' scenario.
Lastly, our work confirms that additional context from version history can play a role in improving code representation for code clone detection and code classification. Future work will investigate the potential improvement for other software engineering tasks.

\section{Research Agenda}

Our preliminary research produces promising outcomes. This section outlines our research agenda to explore different approaches to incorporate version history (and other additional contexts) for better source code representation. 


\textbf{Software Engineering Artefacts.} Recent papers claim that additional artefacts are essential to support software developers and deep learning models in comprehending source code~\cite{tian2022adding}. While mining source code repositories to extract version history data, we can collect different types of artefacts and experiment to encode them into source code representation for downstream tasks. These additional contexts may vary in forms, e.g. natural language (commit messages), graphs (call hierarchy), timestamp (commit date-time), or numerical data (number of days, number of versions)~\cite{thongtanunam2013mining}. 


While mining additional contexts, we encounter challenges like inconsistent availability and imbalances in artefacts. For instance, in the SeSaMe dataset, 80\% of methods have only one or two versions, yet some have over 200, often with minor differences. This disparity can introduce noise and computational inefficiencies in deep learning, underscoring the need for further analysis on how to best encode version history context to code representation.


\textbf{Aggregation and Underlying Models.} Our preliminary research suggests a need for varied aggregation methods for different contexts in code representation, particularly as our current method of concatenating code versions faces limitations with long-sequence data in ASTNN and CodeBERT. ASTNN is constructed on RNN and GRU, which has a long-term dependency problem~\cite{zhao2020rnn, greaves2019statistical}. The algorithm allows the model to learn and connect the previous information to the present task; however, when the information is too long (too many versions), it may lose the connection between the present task and the previous nodes. Besides, CodeBERT was pre-trained to handle input with a maximum length of 512 tokens~\cite{nguyen2023multi}; the version history with longer text length will be truncated.

To address the challenge of multiple code versions, potential solutions could be 1) new algorithms that can encode versions selectively or 2) other neural networks, such as Graph Transformer~\cite{dwivedi2020generalization}, which can handle long-term dependencies and hierarchical information more effectively. We also plan to evaluate the impact of various aggregation techniques, including domain-specific aggregation and general-purpose pooling, for deep learning tasks~\cite{gholamalinezhad2020pooling}. With this research agenda, we aim to recommend how to best use source code and additional context.  


\textbf{Data Availability.}
All the materials produced from this study are available on GitHub~\cite{package}

\textbf{Acknowledgment.}
Patanamon Thongtanunam was supported by the Australian Research Council's Discovery Early Career Researcher Award (DECRA) funding scheme (DE210101091).



\newpage

\bibliographystyle{ACM-Reference-Format}
\bibliography{paper}



\appendix

\end{document}